# Where do Germany's electricity imports come from?

Mirko Schäfer, Tiernan Buckley, Anke Weidlich
INATECH
University of Freiburg
Freiburg, Germany
mirko.schaefer@inatech.uni-freiburg.de

Frank Boerman
TenneT TSO B.V.
Arnhem, The Netherlands

*Abstract*— In 2023, Germany's electricity trade balance shifted from net exports to net imports for the first time since 2002, resulting in an increasing discussion of these imports in the public debate. This study discusses different data driven approaches for the analysis of Germany's cross-border trade, with a focus on the methodological challenges to determine the origin of imported electricity within the framework of European electricity market coupling. While scheduled commercial flows from ENTSO-E are often used as indicators, generally these do not correspond to bilateral exchanges between different market actors. In particular, for day-ahead market coupling only net positions have an economically reasonable interpretation, and scheduled commercial exchanges are defined through ex-post algorithmic calculations. Any measure of the origin of electricity imports thus depends on some underlying interpretation and corresponding method, ranging from local flow patterns to correlations in net positions. To illustrate this dependence on methodological choices, we compare different approaches to determine the origin of electricity imports for hourly European power system data for 2024.

## I. Introduction

The transformation of the German electricity system as part of the "Energiewende" is analyzed and discussed based on a variety of data sources, including generation and load data, prices on different markets, redispatch costs and further indicators. One topic which receives considerable attention also in the public sphere is the increasing amount of electricity imports of Germany. For the last two decades, Germany mostly had been a net exporter of electricity, but since 2023 shifted to being a net importer, with annual imports of 77.2 TWh facing exports of 48.9 TWh according to a report by Agora Energiewende [1]. Given that Germany has phased out its nuclear power plants and is in the process of phasing out its electricity generation from coal, this increase in imports has been discussed for instance in the context of energy security, system and consumer costs, or with respect to the type and origin of electricity imports. The annual trade balance of electricity imports in Germany still does only amount to about 5% of its total power consumption and is rather the result of efficient electricity markets than a sign of critical dependencies [2] [3]. But given the increasing interest in the topic, and the challenges in correctly interpreting various data sources of coupled electricity markets, in this contribution we focus on this question - how much electricity does Germany import, and where do these imports come from?

Most analysis of electricity imports is based on data from the European Network of Transmission System Operators for electricity, ENTSO-E. Through the ENTSO-E Transparency Page [4], hourly values for scheduled commercial exchanges between physically connected bidding zones are published. These values are also republished by the service SMARD [5] from the German Federal Network Agency ("Bundesnetzagentur") and by services like Agorameter [6] from Agora Energiewende or Energy-Charts [7] from Fraunhofer ISE. The name and usage of this data suggests that these flows represent underlying bilateral trade relations between electrically neighboring countries, allowing conclusions about how much electricity has been imported from a specific country in a specific market time unit. However, such an interpretation is misleading.

For the day-ahead market, the published scheduled commercial exchanges are calculated algorithmically ex-post for scheduling purposes and as an analytical instrument, but do not directly represent underlying bilateral trades or necessarily take into account grid restrictions. The calculation takes place after the day-ahead market result, published as scheduled commercial flows day-ahead [8]. In the single day-ahead market coupling (SDAC), an optimum of exchange relationships is calculated across all participating countries, with transmission constraints integrated as restrictions on the set of allowed zonal net import/export values. This Europe-wide cost-efficient market result determines net exchange positions for each market zone to optimize overall welfare [9]. Accordingly, only these zonal net exchange positions have a direct economic meaning in the market data.

In the intraday market, continuous trading between market participants takes place, with trades between different bidding zones algorithmically given a certain path in the system [10]. The superposition of trade paths from the single intraday coupling (SIDC) and the scheduled commercial flows from the day-ahead market result are published by ENTSO-E as scheduled commercial flows total [10]. This dataset is most commonly used for electricity import and export analysis. Consequently, these commercial flows cannot be directly used for determining the origin of imports: In the day-ahead market, the market result is based on net imports and net exports, not bilateral trades, and in the intraday market, flows represent only links in different paths, without information about the endpoints.

Nevertheless, the question regarding the origin of electricity imports will persist, and for some applications need to be addressed using transparent methods. To illustrate the dependence of total imports and their shares with respect to origin or generation type, in the following we introduce some methods and their underlying interpretation, and show results based on hourly power system data for 2024. Subsequently, we also discuss how different data sources for zonal net positions represent different stages in the market coupling process, which needs to be considered when interpreting the data for analysis.

## II. DATA AND METHODS

For our analysis we use power system data published through the ENTSO-E Transparency Platform [4]. We include time series for generation per type, storage charging and discharging, cross-border physical flows, and scheduled commercial flows (day-ahead and total) and SDAC net positions for bidding zones in the ENTSO-E area. We do not correct different underlying coverages or reporting procedures in the published data [11]. Additionally, we include generation and load data from or to the UK. If necessary, time series are converted to hourly resolution. All calculations are done for bidding zones, with bidding zones in the same country aggregated in final results when considering the origin of imports.

In the following we define different methods to calculate the composition of imports with respect to their origin. Detailed mathematical formulations and resulting time series for all measures 2021 to 2024 can be found in the supplementary material [12]. Once the import quantity from a specific origin is determined for a given hour, this value is multiplied by the generation mix in the originating region at this time, yielding the amount of imports from a certain location and generation type. Aggregating over all regions then determines the import quantity associated with a specific generation type for a given hour.

Additionally, we consider different values of net positions of the Germany-Luxembourg bidding zone, based on SDAC net positions, scheduled commercial flows day-ahead and total, and cross-border physical flows. This dataset is available in [13].

### A. Economic interpretation

Although we argue in the introduction that scheduled commercial flows (total) between Germany and a physical neighbor cannot be directly interpreted as bilateral trades, we include the import measure based on this approach for comparison. Accordingly, imports based on *commercial flows total (CFT)* are thus simply given by the original value as published by ENTSO-E. This approach, for instance, has been used in [1] to determine the import mix of Germany in 2024.

Since in the continuous intraday market different trades can take place with paths in both directions over a physical connection, the scheduled commercial flows total often contain non-zero values in both directions. If these values are directly used as in *CFT*, this corresponds to both imports from and exports to a specific neighbor in the same hour. We define imports based on *netted commercial flows total (netted CFT)* as the net scheduled commercial flow with a physical neighbor in each hour. Accordingly, total imports based on this latter interpretation are always smaller or equal to imports based directly on commercial flows total, due to export flows being subtracted from import flows over a connection. This definition, for instance, is used in the analysis presented on the service website SMARD by the German Network Federal Agency [14]. It should be noted that import balances, i.e. the difference between total imports and exports, are identical for *CFT* and *netted CFT*. Differences between the results for import balances in [14] and [1] result from considering Germany as a country in [1] and the German-Luxembourg bidding zone in [14].

In both definitions the origin of imports is always limited to electrically neighboring zones of Germany. To account for the coupled electricity markets over the entire ENTSO-E region, we include two more measures based on the entire import/export flows in the system. For imports based on *pooled netted commercial flows total (pooled netted CFT)*, we first net over import/export values on each connection separately, yielding a pattern of either import or export flows between bidding zones. We then sum up all export values over all connections in the ENTSO-E area to determine the overall export mix for a specific hour. This export mix defines the import mix for all imports in the system. The total imports are thus identical to the ones derived from *netted CFT*, but the composition is now given by a global import mix instead of the local mix assumed in the previous definitions.

Imports based on *pooled net commercial imports total (pooled net CFT)* apply a further netting operation. Scheduled commercial flows total associated with a bidding zone are aggregated over all connections to its electrically neighboring zones, yielding the net commercial import total for a given hour. Analogously to the last measure, all net exports from bidding zones are then aggregated to yield the system-wide export mix, which is applied to the net imports of regions as the overall import mix. Since in this interpretation net positions of bidding zones instead of net positions over individual connections are considered, the total imports are always smaller or equal to the ones based on *net CFT*.

### B. Physical interpretation

The measures defined in the last paragraphs are all based on scheduled commercial exchanges total, following a potential market-based interpretation of trades between different parties across bidding zones. Alternatively, one can follow a physical approach to try to estimate the origin of power flows in the transmission grid, represented by cross-border physical flows as published by ENTSO-E. Similar to the *pooled net CFT* presented above, we define *pooled net physical imports (pooled net phys.)* based on the hourly net positions of bidding zones with respect to cross-border physical flows. All net exports are aggregated, and the resulting physical export mix applied to all net imports of the various regions. This approach approximates the power grid as a copper plate, distributing net exports equally to all net importers.

Flow tracing is an intuitive approach to define the origin of electricity imports based on physical flow patterns, considering the network structure of the underlying electricity grid. Two different approaches need to be distinguished [15]. In *aggregated coupling flow tracing (AC flow tracing)*, net exports of a region are mixed with ingoing power flows from neighbors. This mix is then exported over all outgoing power flows. Physical net imports are thus composed of net exports from other regions, with exporters more closely located in the power flow pattern providing a higher share. This approach, for instance, is used for the calculation of consumption-based emission intensity time series of German federal states in the service CO2Map.de [16]. Since for *AC flow tracing* physical net imports are used, the total imports correspond to the ones obtained from *pooled net physical imports*. An alternative approach denoted as *direct coupling flow tracing (DC flow tracing)* also needs to consider generation and load time series inside the individual regions. Here, we use per type generation and storage discharging as published by ENTSO-E, and define the load as the difference between physical imports plus generation including storage discharge and physical exports, following the approach in [17] also used by the service Electricity Maps [18]. The resulting values often differ from the actual load time series as published by ENTSO-E due to different approximations and limited coverage of the close to real-time power system data reported by transmission system operators. For *DC flow tracing*, incoming physical power flows are mixed with local generation including storage discharge. This mix is then attributed both to the local electricity demand and the outgoing power flows. Accordingly, in this interpretation imports are also attributed to net exporters due to transient flows. Total imports thus are higher than total imports based on net imports when applying *AC flow tracing* or *pooled net phys.*

## C. Net positions

Zonal net positions are relevant for economic analysis, for instance with respect to price spreads or more generally price formation in the coupled European electricity markets. However, if information about net positions is integrated in economic analysis, different data sources for different market situations must be distinguished. In the following, net flows always refer to the net position as calculated by aggregating all ingoing and outgoing flows over connections to electrically neighboring regions.

The *SDAC net positions* as published by ENTSO-E (Implicit Allocations – net positions) refer to the net position of a bidding zone after single day-ahead market coupling. These net positions result from the market coupling algorithm (EUPHEMIA) for the day-ahead market. *Net commercial flows day-head (net CFDA)* represent the net position after the day-ahead market, taking into consideration also cross-border trade with markets not included in the SDAC, most notably Switzerland and the UK. *Net commercial flows total (net CFT)* include further transactions in the single intraday market coupling (SIDC). These net positions still can be different from the final *net (cross-border) physical flows (net phys.)* due to further balancing measures by the transmission system operators. These differences should be considered for the analysis of specific market situations. For instance, net positions from *net commercial flows total* are often interpreted based on mechanisms in the single day-ahead market coupling, thus disregarding that these net positions can include significant influences from commercial exchanges outside the SDAC with Switzerland, or as part of the SIDC.

### III. RESULTS

#### A. Imports

Figs. 1 and 2 display electricity imports of the German-Luxembourg bidding zone in 2024 analyzed according to the different measures introduced in Secs. II.A and II.B. The highest import value of 77.2 TWh is obtained for the sum over all incoming commercial flows, where no netting is performed. This reduces to 66.9 TWh if hourly commercial flows are netted for each individual connection. In the case of considering net imports, i.e. netting over all ingoing and outgoing flows, total imports reduce to 43.3 TWh for commercial flows (total) and 40.8 TWh for cross-border physical flows, respectively. Imports based on flow tracing with direct coupling amount to 62.3 TWh, since in this measure local generation is mixed with physical inflow, such that a region potentially is assumed to import electricity even if it is a net exporter [15].

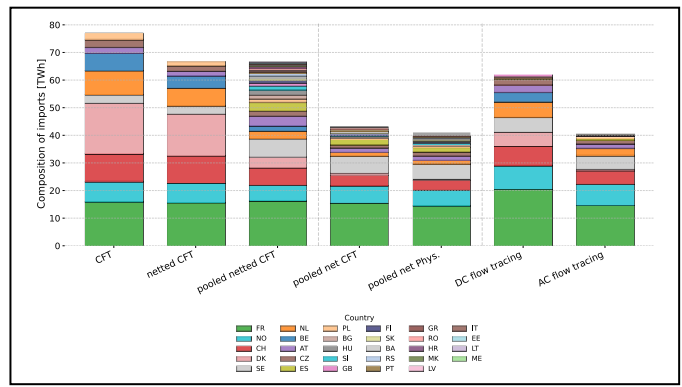

Figure 1. Composition of imports of the German-Luxembourg bidding zone in 2024 with respect to country of origin

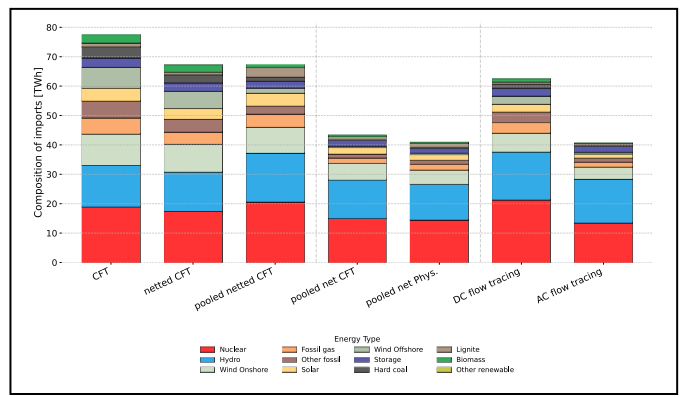

Figure 2. Composition of imports of the German-Luxembourg bidding zone in 2024 with respect to generation type

Fig. 1 shows that for all measures, France is a main origin of imports with import values between 14.6 TWh (*pooled net phys.*) and 20.6 TWh (*DC flow tracing*). This corresponds to its role as a major net exporter of electricity in the European market, and its geographic proximity to Germany. Some countries with larger contributions in all measures are Switzerland (3.5 TWh to 10.2 TWh), Norway (5.7 TWh to 8.6 TWh) and Sweden (3.0 TWh to 6.6 TWh). For the Netherlands and in particular for Denmark, the different measures show considerably varying results. Both countries show higher shares (up to 18.2 TWh for Denmark and 8.8 TWh for the Netherlands) for measures which are either based directly on commercial flows (*CFT*) or derived from flow tracing with direct coupling (*DC flow tracing*). These measures have a higher emphasis on local flow patterns, less on net imports in the overall system. Based on *pooled net CFT*, these shares reduce to 0.35 TWh and 1.4 TWh, respectively, given that this measure only considers hourly net positions and is independent of geographic locations. The reverse case is obtained for Sweden, which has a higher share of up to 15% for measures based on net positions, reduced to 4% for the measure based directly on commercial flows total (*CFT*).

Fig. 2 shows the imports according to generation type, based on multiplication of hourly import shares with the generation mix at each location. The largest contributions in most measures are associated with nuclear (between 14.6 TWh and 21.5 TWh) and hydro power (between 12.1 TWh and 16.6 TWh), due to their role in the net exporting countries France, Switzerland, Sweden and Norway. Renewable generation from wind (onshore and offshore) and solar shows import values between 7.5 TWh and 22.2 TWh.

*B. Net positions*

Tab. 1 shows annually aggregated values and some specific hourly values for net positions of the German-Luxembourg bidding zone based on *SDAC* net positions, net commercial flows day-ahead (n*et CFDA*), net commercial flows total (n*et CFT*), and net cross-border physical flows (n*et phys.*). Additional, net load and residual load as published by SMARD are shown [5]. Whereas total export values do not significantly differ between these measures, total import values based on SDAC net positions are 6 to 7 TWh lower than values from the other measures. This difference is related to cross-border exchanges with Switzerland, which are not considered in the SDAC. Further discrepancies become apparent in the analysis of specific hourly values. Tab. 1 includes net and residual load and hourly values for the four hours with the highest import net positions for each measure, respectively. The results show that these net positions vary in different ways, depending on the specific situation. For example, the hour with the highest negative net position in SDAC shows significant renewable generation, as represented by the lower residual load. In this hour, imports then have been reduced in the SIDC. The hour with the highest negative net position based on net commercial flows total is present for a situation with very low renewable generation. Here, exchanges day-ahead outside the SDAC (notably Switzerland) and in SIDC lead to increases in net imports of approx. 5 GWh for this hour.

Table 1: AGGREGATED AND SPECIFIC HOURLY VALUES FOR NET POSITIONS BASED ON SINGLE DAY-AHEAD COUPLING, NET COMMERCIAL FLOWS DAY-HEAD, NET COMMERCIAL FLOWS TOTAL, AND NET CROSS-BORDER PHYSICAL FLOWS

|  | SDAC | net CFDA | net CFT | net phys. |
|---|---|---|---|---|
| **2024** | Net load: 468.9 TWh | | Residual load: 267.4 TWh | |
| Exports [TWh] | 11.8 | 12.7 | 11.5 | 12.0 |
| Imports [TWh] | 35.7 | 43.0 | 43.3 | 40.8 |
| Balance [TWh] | 23.9 | 30.3 | 31.9 | 28.9 |
| 2024-03-27 15:00 CET | Net load: 60.9 GWh | | Residual load: 41.0 GWh | |
| Net position [GWh] | -14.2 | -12.9 | -12.9 | -12.5 |
| 2024-04-10 19:00 CET | Net load: 59.7 GWh | | Residual load: 50.0 GWh | |
| Net position [GWh] | -14.0 | -18.0 | -16.8 | -15.5 |
| 2024-12-12 8:00 CET | Net load: 66.9 GWh | | Residual load: 65.7 GWh | |
| Net position [GWh] | -12.4 | -15.8 | -17.7 | -17.4 |
| 2024-08-18 11:00 CET | Net load: 46.8 GWh | | Residual load: 30.4 GWh | |
| Net position [GWh] | -13.6 | -16.4 | -17.2 | -17.6 |

IV. CONCLUSION

In the coupled European markets, the origin of electricity imports cannot be clearly defined, and for economic analysis mostly does not have a direct economic meaning. In single day-ahead market coupling, only net positions are economically relevant. In other markets, bilateral trade relationships implying a specific origin of an import might be present but are not represented by the scheduled commercial flows often used for analysis, which are calculated algorithmically after SDAC or throughout SIDC market clearing. Accordingly, any measure regarding the origin and type of electricity imports is an ex-post analytical tool, based on some underlying interpretation and with its results dependent on methodological choices, both regarding the composition, but also the total amount of imports.

We thus recommend using net positions for economic analysis, considering that different measures relate to different market situations. If information about the sum and origin of imports is needed for specific applications, calculation methods and data sources must be clearly defined to prevent inconsistencies and misconceptions in the analysis of cross-border trades in the coupled European electricity markets.

ACKNOWLEDGMENT

We thank Leonhard Probst, Christoph Maurer and Pierre Segonne for helpful discussions, and Ramiz Qussous for his support with the preparation of the final figures.